%% file: eom11_8083_48_burger_arxiv.tex
\documentclass[a4paper]{spie}  
\usepackage[]{graphicx}
\usepackage{amssymb,amsmath,psfrag,url}

\title{Rigorous Simulations of 3D Patterns on \\ Extreme Ultraviolet Lithography Masks}

\author{
Sven Burger,\supit{\,ab}
Lin Zschiedrich,\supit{\,b}
Jan Pomplun,\supit{\,b}
Frank Schmidt\supit{\,ab}
\skiplinehalf
\supit{a}
Zuse Institute Berlin (ZIB),
Takustra{\ss}e 7,
D\,--\,14\,195 Berlin,
Germany
\smallskip\\
\supit{b}
JCMwave GmbH,
Bolivarallee 22, 
D\,--\,14\,050 Berlin,
Germany
}
\authorinfo{
Corresponding author: S. Burger\\
URL: http://www.zib.de/en/numerik/computational-nano-optics.html\\
URL: http://www.jcmwave.com\\
Email: burger@zib.de
}

\begin{document}
\maketitle
\noindent
This paper will be published in Proc.~SPIE Vol. {\bf 8083}
(2011) 80831B 
({\it Modeling Aspects in Optical Metrology III; B.~Bodermann, H.~Bosse, R.~M.~Silver, Editors}, 
DOI: 10.1117/12.889831), 
and is made available 
as an electronic preprint with permission of SPIE. 
One print or electronic copy may be made for personal use only. 
Systematic or multiple reproduction, distribution to multiple 
locations via electronic or other means, duplication of any 
material in this paper for a fee or for commercial purposes, 
or modification of the content of the paper are prohibited.

\begin{abstract}
Simulations of light scattering off an extreme ultraviolet lithography mask with a 2D-periodic absorber pattern are presented. 
In a detailed convergence study it is shown that accurate results can be attained for relatively large 
3D computational domains and  in the presence of sidewall-angles and corner-roundings. 
\end{abstract}


\section{Introduction}

Extreme ultraviolet (EUV) lithography at a wavelength of about 13\,nm is expected 
to replace DUV photo\-litho\-graphy for manufacturing features on integrated circuits 
with critical dimensions as small as 22\,nm or beyond. 
Rigorous simulations of light propagation through photomasks are an essential 
component in optical metrology of such structures.  
Rigorous simulations are also used for optimizing feature geometries on masks for 
improving lithographic process stability and for resolution enhancement in printing of sub-wavelength 
features (computational lithography). 

\begin{figure}[t]
\begin{center}
  \includegraphics[width=.6\textwidth]{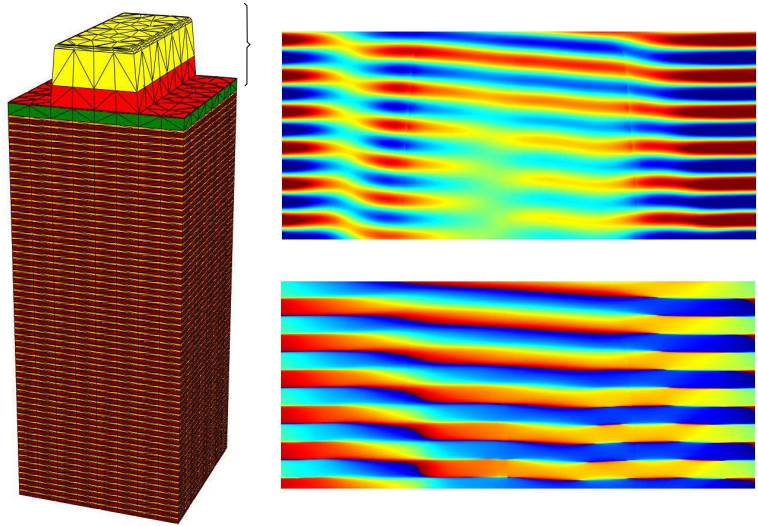}
  \caption{
Left: 3D computational domain: Absorber structure on a multi-layer mirror.
Right (top/bottom): Real part/phase of the dominant field component of a 3D vectorial electric field. 
Obtained from an export of the 3D field to a 2D cross section 
through the upper part (as indicated) of the computational domain, pseudo-color representations.
}
\label{fig_block_3d}
\end{center}
\end{figure}

In deep ultraviolet (DUV, wavelength of about 193\,nm) lithography and metrology 
simulations a main challenge consists in accurate 
resolution of light fields in the presence of complex 3D absorbing structures of high 
refractive index-contrasts. 
In the EUV regime available materials exhibit far lower refractive index-contrasts. 
On the one hand this simplifies computations because high field enhancements and field singularities 
do not occur. 
On the other hand the limits to available optical materials in the EUV regime
leads to  additional challenges for rigorous simulations: 
(i)~Computational domain sizes increase due to the fact that absorber structures need 
larger volumes (relative to the cubic illumination wavelength). 
(ii)~Deviations from ideal geometries like sidewall-angles have a larger effect on the 
diffraction spectra. 
(iii)~EUV masks are typically mounted on multi-layer mirrors with a high number of single layers. 
This again increases 3D computational domain size and complexity. 

We have developed a general finite-element (FEM) Maxwell solver which also allows to address 
3D EUV simulation tasks. 
The solver incorporates higher-order edge-elements, domain-decomposition methods 
and fast solution algorithms for solving time-harmonic Maxwell's equations in 
various problem formulations (e.g., resonance, scattering type problems)~\cite{Burger2008ipnra}. 
Previously the solver has been used for the study and metrological investigations 
of EUV line masks (1D-periodic patterns)~\cite{Pomplun2006bacus,Scholze2007a,Tezuka2007a,Scholze2008a}.
In this contribution we report on rigorous electromagnetic field simulations of 2D-periodic 
arrays of absorber structures on EUV masks.  
In a convergence study we show that highly  accurate results can be attained.

This paper is structured as follows: 
The investigated mask setup is described in Section~\ref{section_setup}. 
In Section~\ref{sec_convergence} a convergence study is performed: In Section~\ref{sec_reference_1d_line} highly 
accurate results (reference solution) for the scattering response of a line mask are generated using a 2D light scattering 
solver. Consistent solutions are generated using two different approaches: 
(i)~scattering off the full structure, and (ii)~scattering off the absorber structure and the multi-layer mirror 
coupled through a rigorous domain-decomposition approach~\cite{Schaedle_jcp_2007}. 
In Section~\ref{sec_3d_1d_line} the previously obtained accurate results are used as reference solution to 
investigate convergence of the full 3D light scattering solvers. 
In Section~\ref{sec_3d_simulations} the solver is used for simulation of the full 3D problem of a 2D-periodic array of 
absorber structures on an EUV mask. Again, numerical convergence of the obtained results is investigated. 
Numerical simulation results are tabulated in the Appendix.

\section{Investigated setup}
\label{section_setup}

\begin{figure}[t]
\begin{center}
\psfrag{R}{\sffamily $R$}
\psfrag{alpha}{\sffamily $\alpha$}
\psfrag{ha}{\sffamily $h_a$}
\psfrag{hb}{\sffamily $h_b$}
\psfrag{hox}{\sffamily $h_{ox}$}
\psfrag{hcap}{\sffamily $h_{cap}$}
\psfrag{hmo}{\sffamily $h_{mo}$}
\psfrag{hsi}{\sffamily $h_{si}$}
\psfrag{ha}{\sffamily $h_a$}
\psfrag{absorber}{\sffamily absorber}
\psfrag{buffer}{\sffamily buffer}
\psfrag{oxide}{\sffamily oxide}
\psfrag{capping}{\sffamily capping}
\psfrag{multilayer}{\sffamily multi-layer (60 $\times$ 2)}
\psfrag{substrate}{\sffamily substrate}
\psfrag{px}{\sffamily $p_x$}
\psfrag{cd}{\sffamily $CD_{bottom}$}
  \includegraphics[width=.4\textwidth]{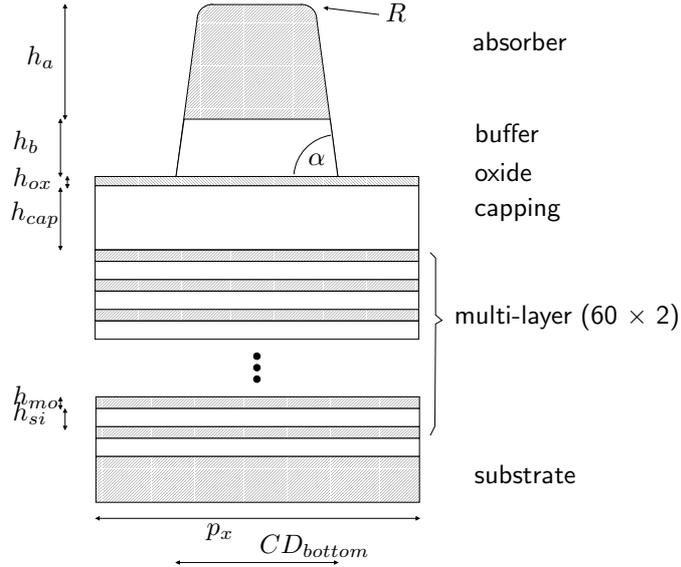}
  \caption{
Schematics of the 2D setup (and of the 2D cross-section through the 3D setup). 
See also Fig.~\ref{field_3d_block} for the definition of the lateral geometry. 
}
\label{schematics_euv}
\end{center}
\end{figure}

The investigated structures consist of an absorber stack on a multi-layer mirror (consisting of a total of 120 layers). 
Figure~\ref{fig_block_3d} shows a 3D mesh of the geometry and a typical electric field distribution in  a part of the stack. 
A schematics of the setup is shown in Figure~\ref{schematics_euv}. 
This study is concerned with numerical properties of the FEM simulation method, therefore only one fixed setting of the 
material and stack parameters is investigated, and only two fixed settings of lateral placement of absorber structures 
on the multi-layer (a 1D-periodic line mask and a 2D-periodic pattern).
The chosen geometrical and material parameters of the physical setting are given in Table~\ref{table_parameters}.
For modeling unpolarized illumination, the near-fields corresponding to illumination with 
S- and P-polarized plane waves at 13.4\,nm vacuum wavelength and at an (in-plane) angle of incidence of 4~degree
are computed, and an incoherent superposition of the fields is performed. 
%
Highly accurate, rigorous numerical simulation of this 3D setup is challenging because the total size of the 3D computational domain is 
about $1.5\times10^7\,\mbox{nm}^3$ which corresponds to about $6,300\,$cubic wavelengths. 

\begin{table}[h]
\begin{center}
\begin{tabular}{|l|l|l|l|l|}
\hline
material & height & n & k & \\
\hline
air & $\inf$ & 1 & 0 & \\
absorber & $h_a = 40\,$nm & 0.93368 & 0.03791 & $\alpha = 85\,$deg, $R=5\,$nm\\
buffer & $h_b = 20\,$nm  & 0.97468 & 0.01261 & $\alpha = 85\,$deg\\
oxide & $h_{ox} = 1.1\,$nm & 0.97468 & 0.01261 & \\
capping & $h_{cap} = 11.5\,$nm & 1.00024 & 0.00182 & \\  
multi-layer (Mo) & $h_{mo} = 2.42\,$nm & 0.92373 & 0.0061 & 60\,layers \\
multi-layer (Si) & $h_{si} = 4.48\,$nm & 1.00024 & 0.00182 &  60\,layers\\
substrate & $\inf$ & 0.97908 & 0 & \\
\hline
\hline
lateral dimensions & $CD_{bottom,x}$ & \multicolumn{3}{l|}{88\,nm}\\
                   & $CD_{bottom,y}$ & \multicolumn{3}{l|}{110\,nm}\\
                   & $p_x$ & \multicolumn{3}{l|}{176\,nm}\\
                   & $p_y$ & \multicolumn{3}{l|}{176\,nm}\\
\hline
\hline
illumination & angle of incidence & \multicolumn{3}{l|}{$\theta_{\mbox{in}} = 4\,$deg}\\
                   & wavelength & \multicolumn{3}{l|}{$\lambda_0 = 13.4\,$nm}\\
\hline
\end{tabular}
\caption{Parameter settings for the EUV
mask simulations (compare Fig.~\ref{schematics_euv})
Mask geometry parameters (layer heights $h_x$, sidewall angle $\alpha$, corner rounding 
radius $R$), material parameters 
(real and imaginary parts of the refractive index, $n$ and $k$), 
illumination parameters (in-plane angle of incidence, vacuum wavelength).
}
\label{table_parameters}
\end{center}
\end{table}

\section{Convergence study}
\label{sec_convergence}
\subsection{Reference solution: 1D line mask}
\label{sec_reference_1d_line}
The aim of this paper is to demonstrate which accuracy of rigorous electromagnetic field simulations 
can be reached for relatively large computational domains. 
For the general 3D scattering problem (2D-periodic absorber pattern) 
as described in Section~\ref{section_setup} no analytical solution exists.  
No alternative results are available which can be used as quasi-exact results in order to quantify numerical errors of a 3D 
simulation result. 
Therefore the problem is analyzed as follows: 
First, results for a linemask setup are generated using a rigorous FEM solver on a 2D computational domain. 
The FEM solver has been compared to other simulation methods, and it also converges with the expected convergence order, 
therefore it can be assumed that the obtained result is quasi-exact. 
Then,  results for the same physical setting (line mask) are generated on a 3D computational domain. 
The quasi-exact result from the 2D setup is used to measure the reached numerical accuracy of the 
3D results. 
It is expected that the accuracy of results of a 2D-periodic pattern with a computational domain of comparable size will 
be similar to the accuracy obtained for the linemask.

\begin{figure}[t]
\begin{center}
  \includegraphics[width=.3\textwidth]{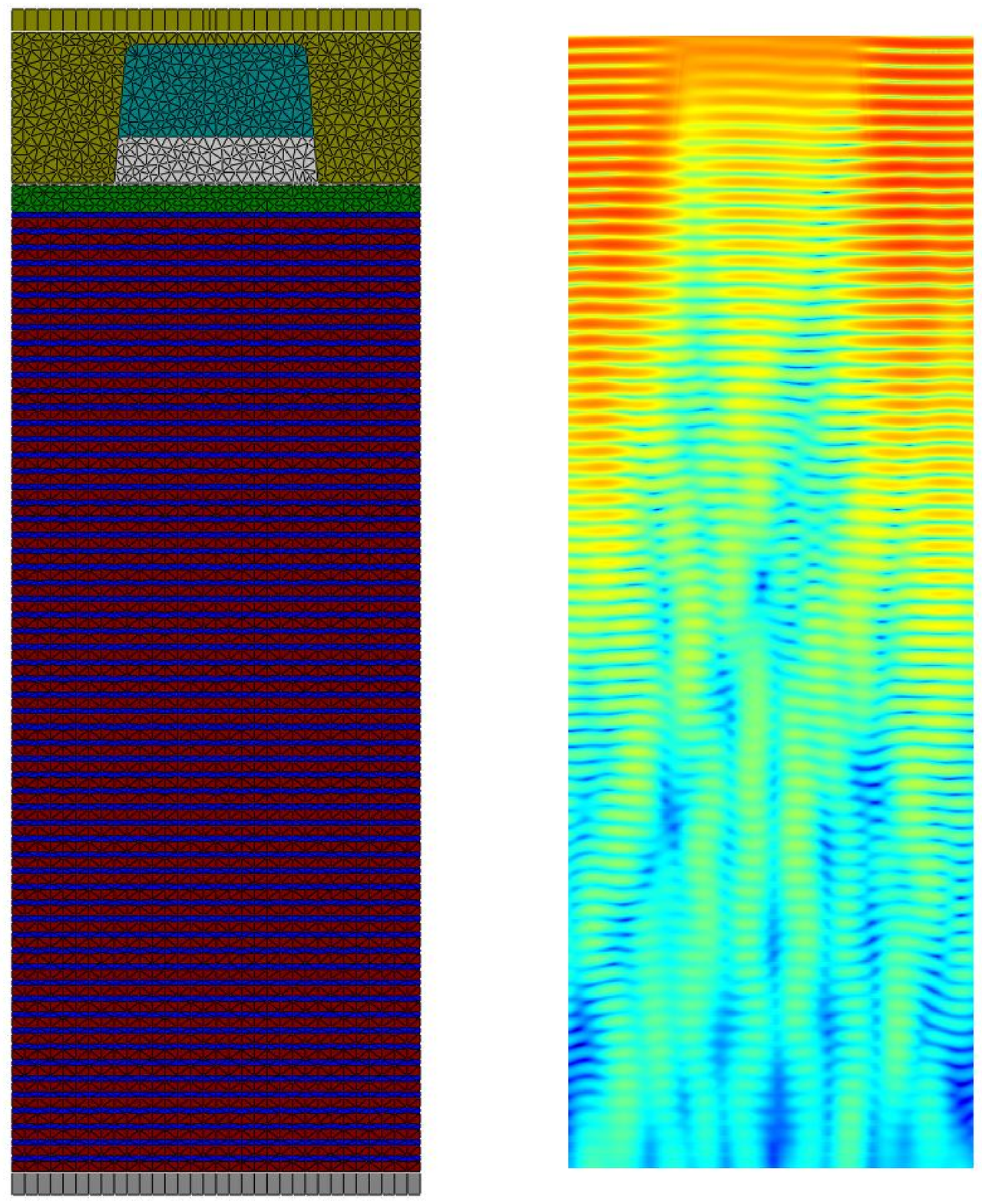}
  \hspace{.03\textwidth}
  \psfrag{N}{\sffamily $N$}
  \psfrag{Delta A}{\sffamily Rel.\,error}
  \includegraphics[width=.3\textwidth]{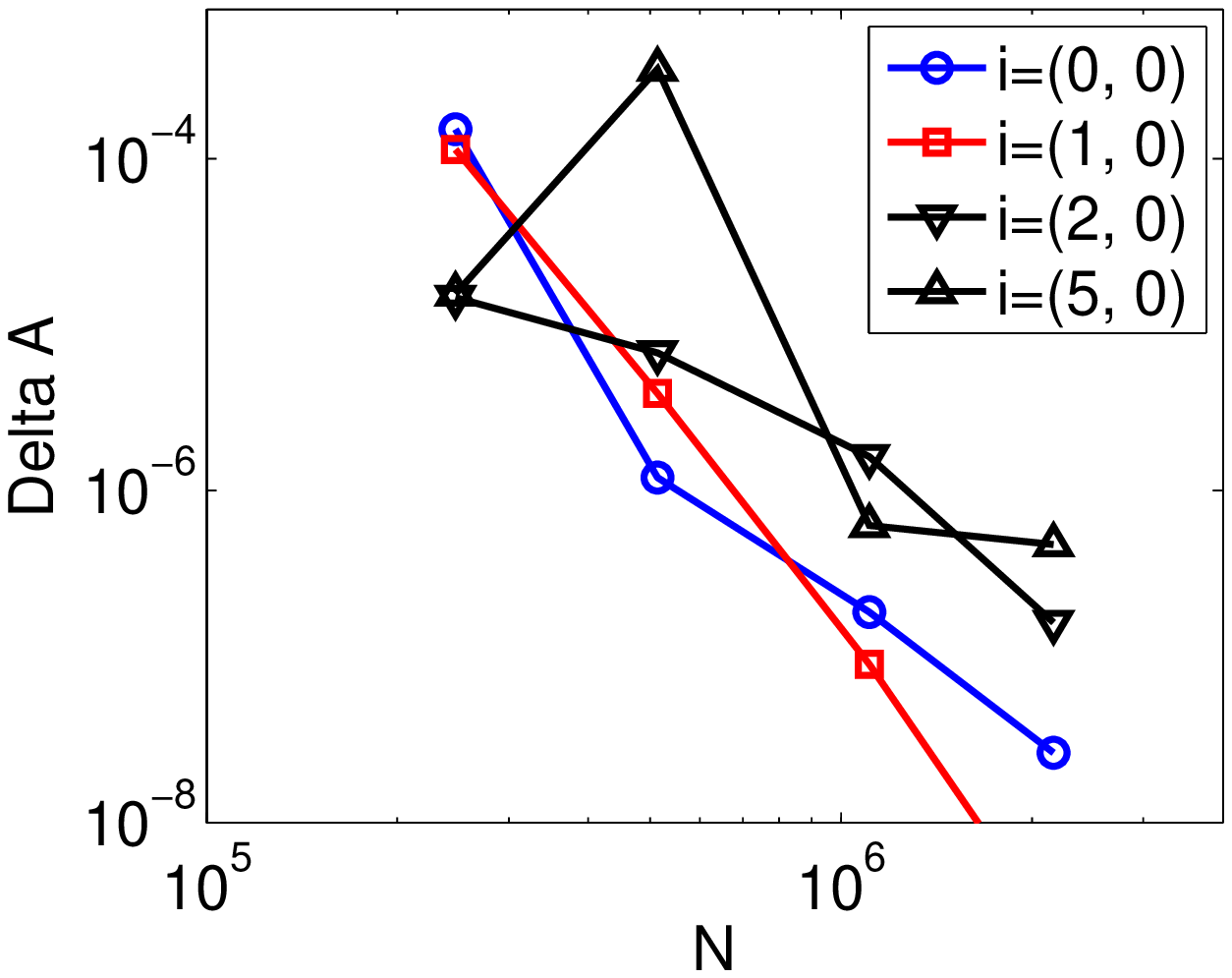}
  \hspace{.03\textwidth}
  \psfrag{Delta P}{\sffamily $\Delta P$}
  \includegraphics[width=.3\textwidth]{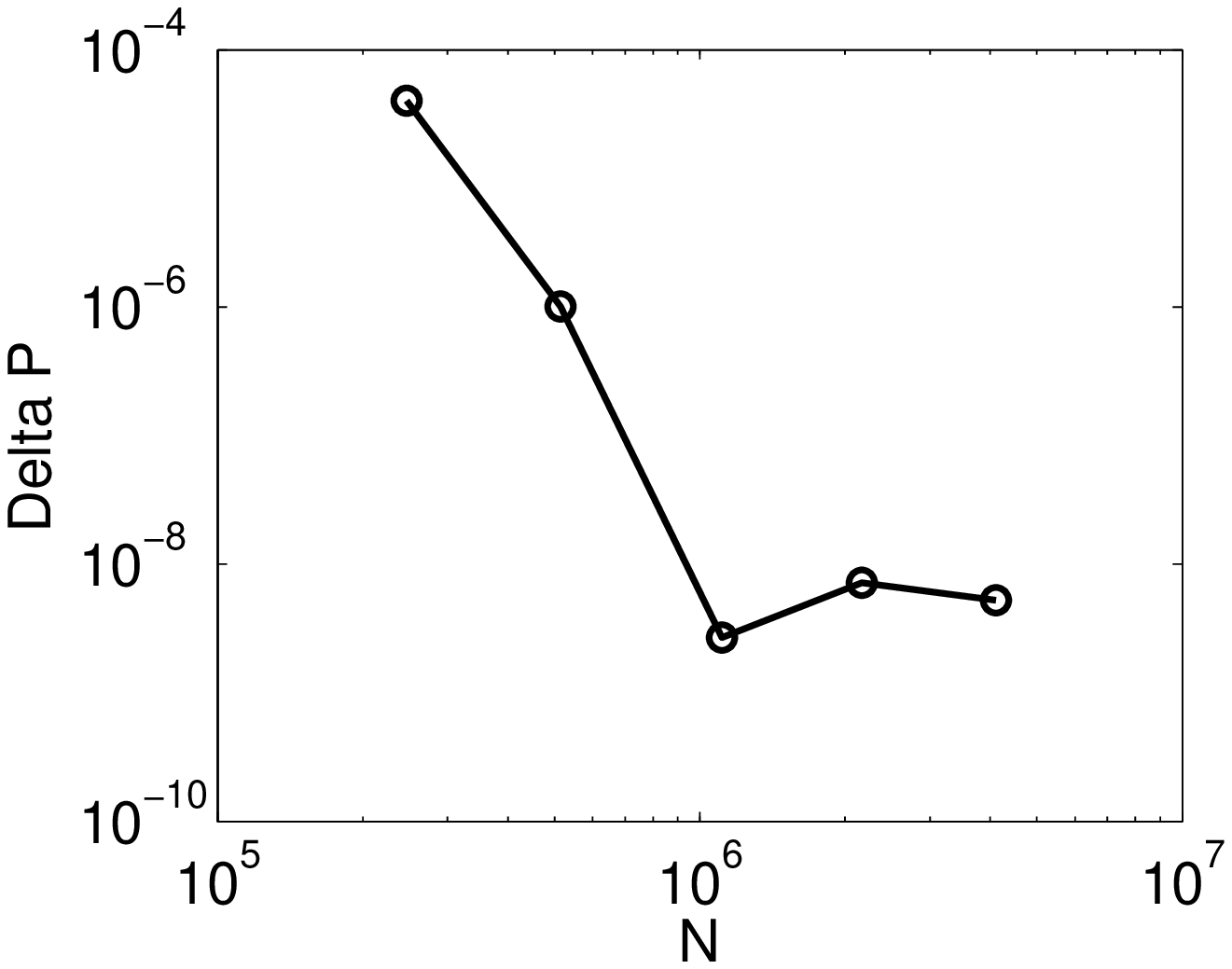}
  \caption{
From left to right: Mesh discretizing the 2D computational domain of a line mask on a 
multi-layer mirror. 
Intensity distribution of the electromagnetic near-field (pseudo-color representation, logarithmic scale).
Relative error of the amplitudes of the zero, first, second and fifth diffraction orders, $\Delta |\vec{A}(\vec{k}_i)|^2$,
relative error of power conservation $\Delta P$,
(cf.~Table~\ref{table_2d_full}).
}
\label{field_2d_full}
\end{center}
\end{figure}

\subsubsection{1D line mask: 2D computational domain}
\label{sec_reference_1d_line_2dfull}
The electromagnetic near-field of the setup as defined in Section~\ref{section_setup} is simulated using a 2D computational 
domain (the structure is invariant in the third dimension). 
Figure~\ref{field_2d_full} (left) shows a typical initial spatial mesh and a graphical representation of a computed near-field 
intensity distribution. 
Various post-processes are used to deduce quantities of interest from the electric near-field $\vec{E}(\vec{x})$: 
\begin{itemize}
\item 
The amplitudes of all propagating, reflected diffraction orders are evaluated by Fourier integration over the 
complex electric field $\vec{E}$ at the upper boundary of the computational domain. 
The propagating Fourier mode $i$ is characterized by the complex amplitude $\vec{A}(\vec{k}_i)$,
the wavevector $\vec{k}_i$ and the diffraction order $(n,m)$. 
$$
\vec{A}(\vec{k}_i)=\frac{1}{L}\int_{a}^{b}\vec{E}(\vec{x})\exp(-i \vec{k}_i\cdot\vec{x}) d\vec{x}
$$
Here, the boundary of the computational domain spans from 
$\vec{x}=\vec{a}$ to $\vec{x}=\vec{b}$.
The Fourier modes are normalized to the width (resp.~area) of the respective computational domain 
boundary, $L=\int_{a}^{b} d\vec{x}$.
The reflected power is then given by  $P_{\mbox{ref}} =\sum_i\cos(\theta_i)|\vec{A}(\vec{k}_i)|^2\times L/Z$, 
with  the angle between the surface normal and the wavevector $\vec{k}_i$ of the respective diffraction 
order, $\theta_i$, impedance $Z=\sqrt{\mu/\epsilon}$, permittivity $\epsilon$ and permeability $\mu$.. 
\item
The amplitudes of all transmitted diffraction orders are obtained in the same way, and 
the transmitted power $P_{\mbox{tra}}$ is obtained. 
\item
The electric field energies $W_{el,j}$ in  all  sub-domains of the computational 
domain, ${D_j}$, are computed by field integration over the complex electric field distribution $\vec{E}$:
$$
W_{el,j}=\int_{D_j}w_{el}d\vec{r}
=\int_{D_j}\frac{(\varepsilon_j \vec{E})^{\ast} \cdot \vec{E}}{4} d\vec{r}.
$$
The absorbed power $P_{\mbox{abs}}$ is then given by  $P_{\mbox{abs}} = 2\omega \sum_j \Im(W_{el,j})$, with the angular frequency 
$\omega = 2\pi c_0/\lambda_0$ and vacuum speed of light $c_0$.
\end{itemize}

The incident power  $P_{\mbox{inc}}$ is given by 
$ P_{\mbox{inc}} = 2\cos(\theta_{\mbox{in}})|\vec{A}_{inc,S/P}|^2\times L/Z$, with the amplitude $\vec{A}_{inc,S/P}$ of the incident S- and P-polarized 
plane waves.
Power conservation is expected, i.e., with increasing numerical resolution 
the sum of incoming, outgoing and absorbed power flux should converge to zero,
 $\Delta P = (P_{\mbox{inc}} - P_{\mbox{abs}} -  P_{\mbox{ref}}- P_{\mbox{tra}})/P_{\mbox{inc}}$, 
$\Delta P \rightarrow 0$. 

In scatterometric applications and in computational lithography applications, the quantities of interest are typically 
the amplitudes of the diffraction orders (because these are detected in a scatterometric setup, resp.~because these 
enter the optical imaging system). 
Therefore this study concentrates on convergence of the results with respect to both, power conservation and intensities of some
few diffraction orders. 

Table~\ref{table_2d_full} shows numerical results: Intensities of some exemplary diffraction orders, of reflected power flux and 
of power conservation have been computed from near-fields obtained at different numerical resolutions. 
For computing these fourth-order finite-elements are chosen (polynomial degree of the finite-element ansatzfunctions, defined 
on the patches of the discretized geometry, $p=4$) and solutions on meshes with different refinement are computed (where 
adaptive mesh refinement steered by an automatic, residual-based error-estimator has been choosen as refinement strategy). 
With increasing mesh refinement the number of geometrical patches is increased which leads to a better resolution of the 
computed electromagnetic near-field and the derived quantities, and which leads to increased number of unknowns $N$ of the 
sparse system of equations resulting from the FEM discretization. Very high numerical accuracy is reached for few refinement steps, 
with total numbers of unknowns below one million and with typical computation times on standard computer hardware in the range of 
few minutes. 
The saturation in power conservation error at a very low error level, $\Delta P<10^{-8}$, may be due to floating point precision errors which may come into 
play at this very high accuracy level. 
Parts of this data is also displayed graphically in Figure~\ref{field_2d_full}.
The displayed relative error of the intensity of diffraction order $i$ is defined as 
$\Delta I_i = (|\vec{A}(\vec{k}_i)|^2 - |\vec{A}_{qe}(\vec{k}_i)|^2)/|\vec{A}_{qe}(\vec{k}_i)|^2$, 
where the FEM simulation result at highest numerical resolution is chosen as quasi-exact result, $\vec{A}_{qe}(\vec{k}_i)$.

\subsubsection{1D line mask: 2D Domain decomposition results}
\label{sec_reference_1d_line_2ddd}

A significant part of the computational effort necessary for the computation of the near-field solution as displayed in 
Figure~\ref{field_2d_full} is necessary for computation of the field distribution in the multi-layer stack. 
This region is essentially only a 1D-structured geometry. Therefore wave propagation in this region can be treated 
quasi-analytically or by solving only 1D FEM problems. However, 
the absorber pattern has a higher dimensionality. 
It has been shown that a rigorous domain-decomposition algorithm can use these properties of the problem for 
reducing significantly the computational effort 
(in terms of numbers of unknowns and computation times)~\cite{Schaedle_jcp_2007}.. 
The domain-decomposition algorithm essentially operates by dividing the computational domain as shown schematically in 
Figure~\ref{schematics_euv} into the multi-layer mirror and the absorber structure. These two domains are then coupled via 
the electromagnetic field coupled back and forth between the domains, and convergence is reached by iterative improvement of 
the field approximations. 

\begin{figure}[t]
\begin{center}
  \includegraphics[width=.3\textwidth]{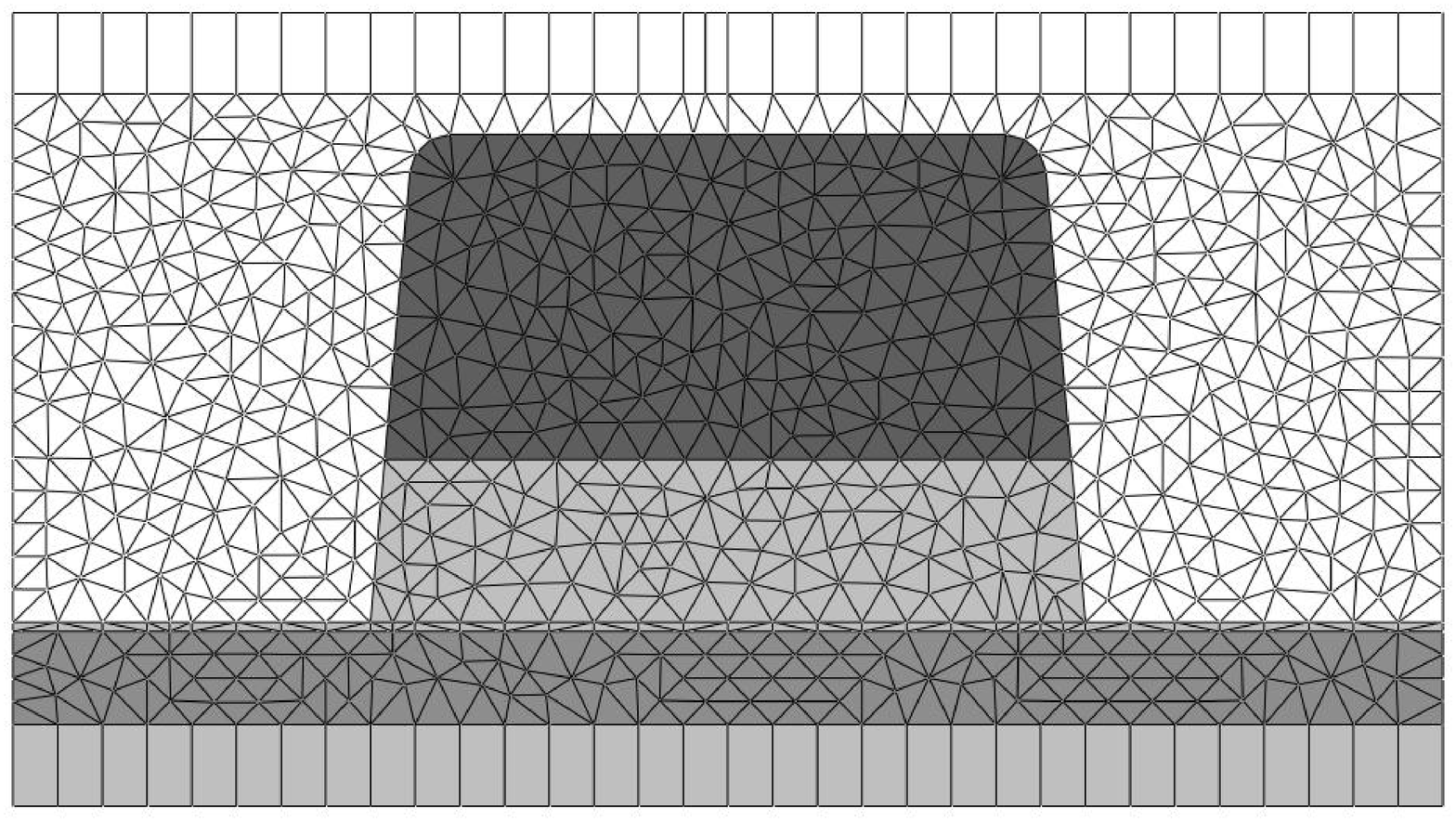}
  \hspace{.03\textwidth}
  \psfrag{N}{\sffamily $N$}
  \psfrag{Delta A}{\sffamily Rel.\,error}
  \includegraphics[width=.3\textwidth]{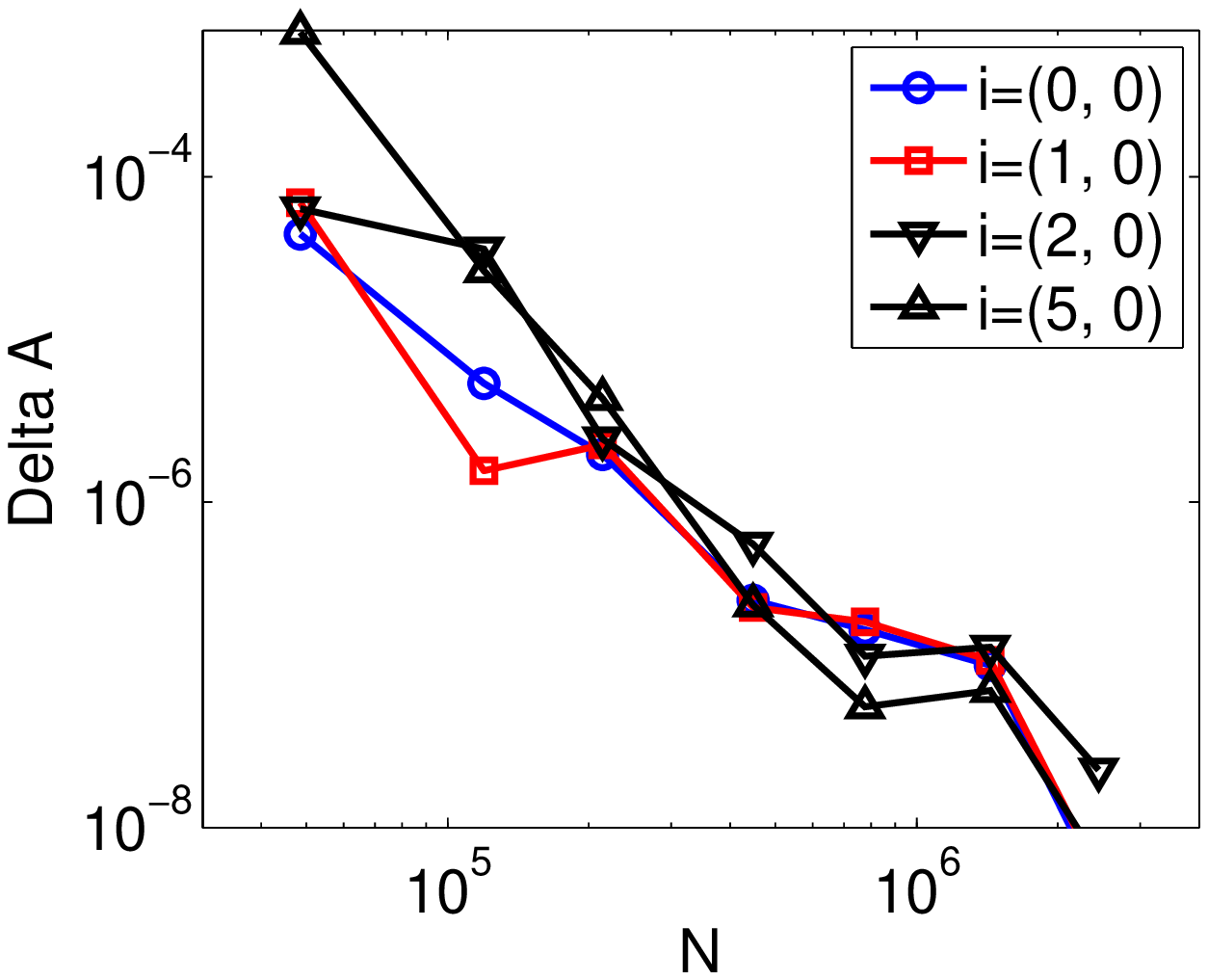}
  \hspace{.03\textwidth}
  \psfrag{Delta P}{\sffamily $\Delta P_{\mbox{ref}}$}
  \includegraphics[width=.3\textwidth]{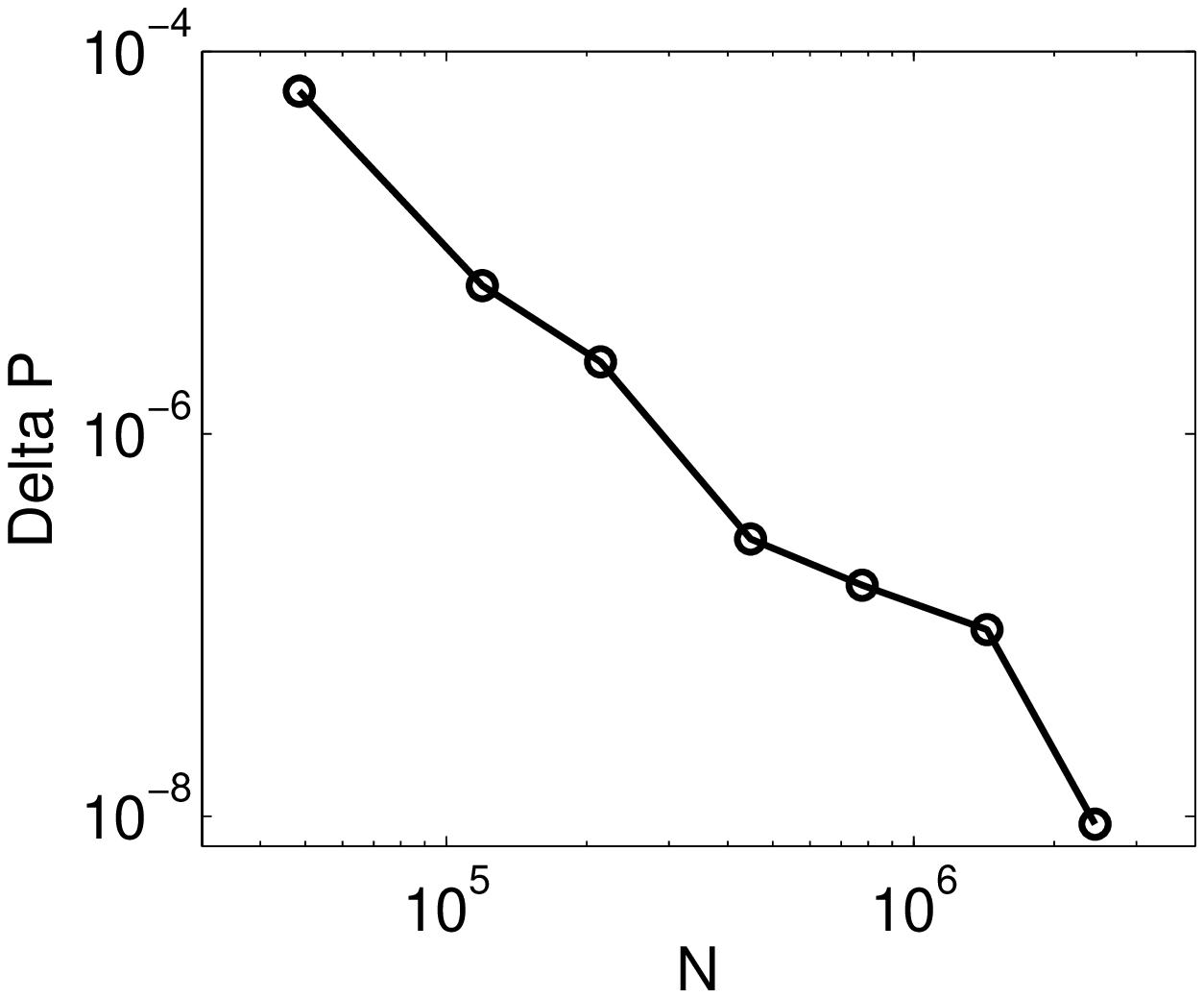}
  \caption{
Results from simulation with the 2D domain-decomposition setup. 
From left to right: 
Mesh discretizing the 2D computational domain of a line mask (reduced computational domain in the 
domain-decomposition approach). 
Relative error of the intensities of the zero, first, second and fifth in-plane diffraction orders, $\Delta |\vec{A}(\vec{k}_i)|^2$,
relative error of reflected power  $\Delta P_{\mbox{ref}}$,
(cf.~Table~\ref{table_2d_dd}).
}
\label{field_2d_dd}
\end{center}
\end{figure}

The domain-decomposition algorithm is applied to the 2D setup, as discussed above. 
Table~\ref{table_2d_dd} shows some of the results (in this case power conservation is not checked, as the absorbed power 
in the multi-stack mirror is not automatically evaluated by the software in the domain-decomposition setup). 
Figure~\ref{field_2d_dd} displays convergence of the diffraction intensities and of the reflected power. As quasi-exact value 
the near-field result with highest numerical resolution has been chosen. 

As can be seen from Table~\ref{table_2d_dd}, the quantitative results agree between the domain-decomposition setup and 
the full 2D setup up to a relative accuracy of about $10^{-8}$ in all investigated quantities. With the domain-decomposition 
setup, very accurate results can be obtained at relatively low computational effort. 

\subsection{3D simulations: 1D line mask}
\label{sec_3d_1d_line}
The main purpose of the previous sections was to compute an accurate reference solution for 
rigorous EMF simulations on a 3D computational domain. In this section the light scattering 
response of the same EUV line-mask is computed using a 3D computational domain. 
The problem is again approached using, first, a full computational domain and, second, a separation 
in multi-layer mirror and absorber using a domain-decomposition solver. 
A 3D problem which cannot be reduced to a 2D setting will be treated in Section~\ref{sec_3d_simulations}.

\subsubsection{1D line mask: Full 3D computational domain}
\label{sec_1d_line_3dfull}
\begin{figure}[t]
\begin{center}
  \includegraphics[width=.15\textwidth]{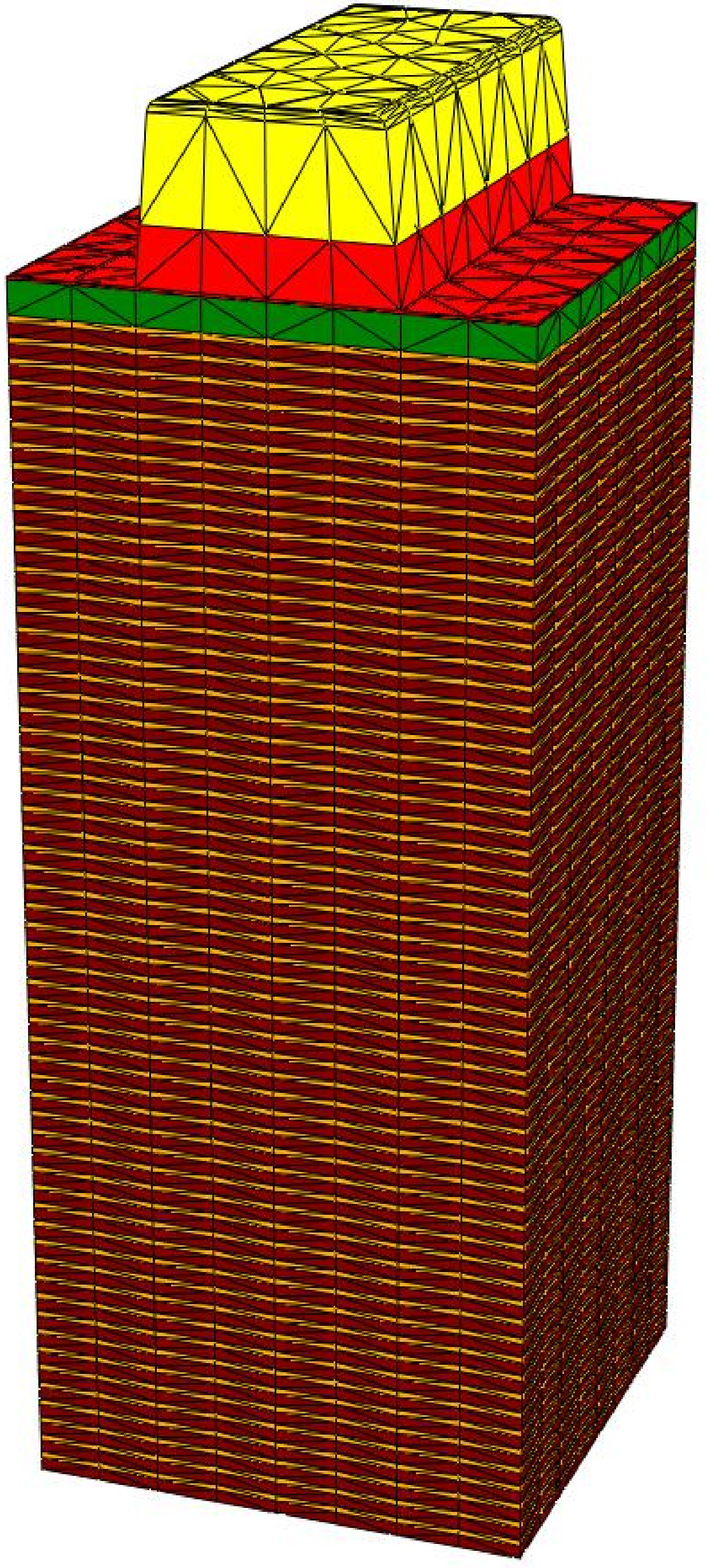}
  \hspace{.03\textwidth}
  \psfrag{N}{\sffamily $N$}
  \psfrag{Delta A}{\sffamily Rel.\,error}
  \includegraphics[width=.3\textwidth]{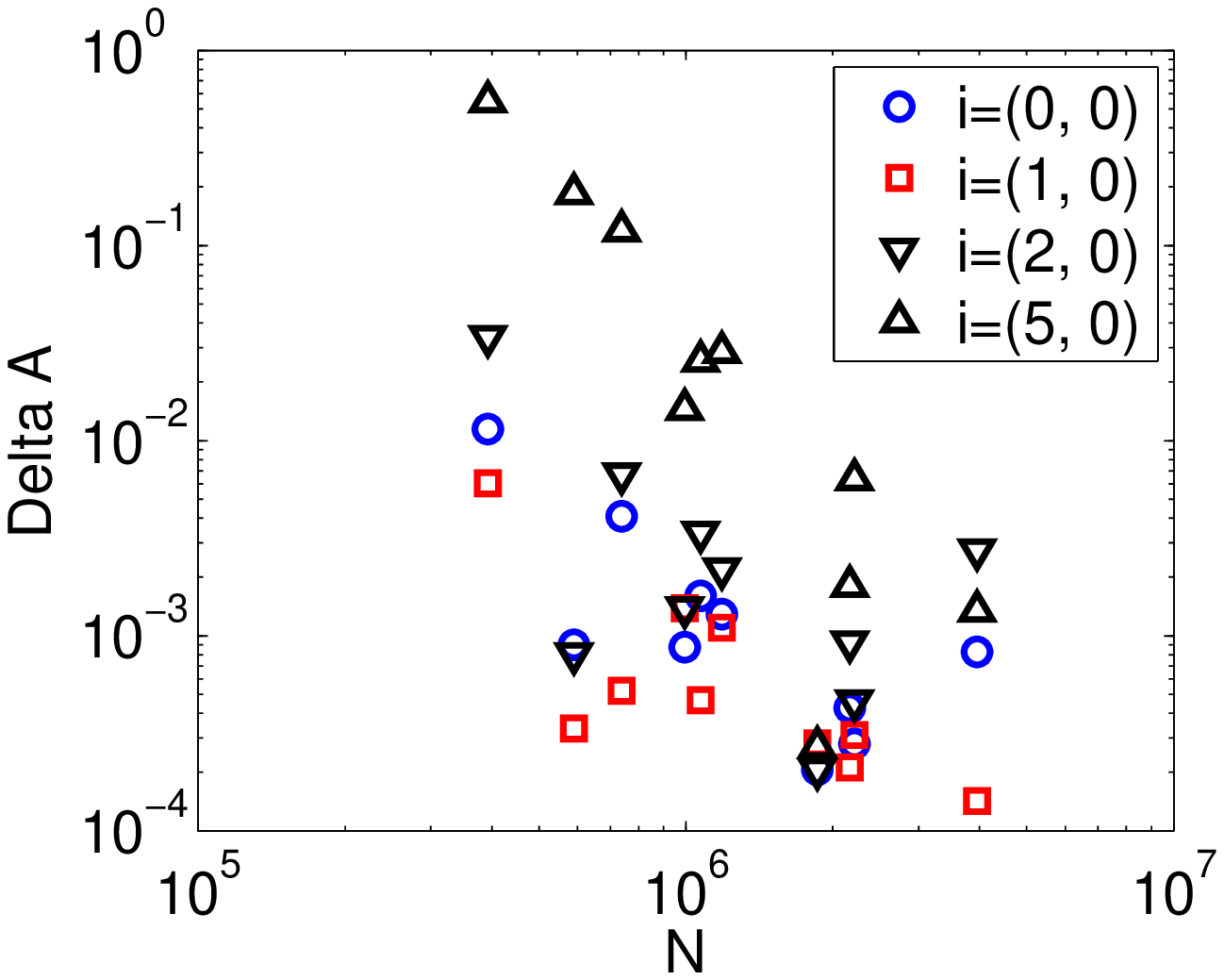}
  \hspace{.03\textwidth}
  \psfrag{Delta P}{\sffamily $\Delta P$}
  \includegraphics[width=.3\textwidth]{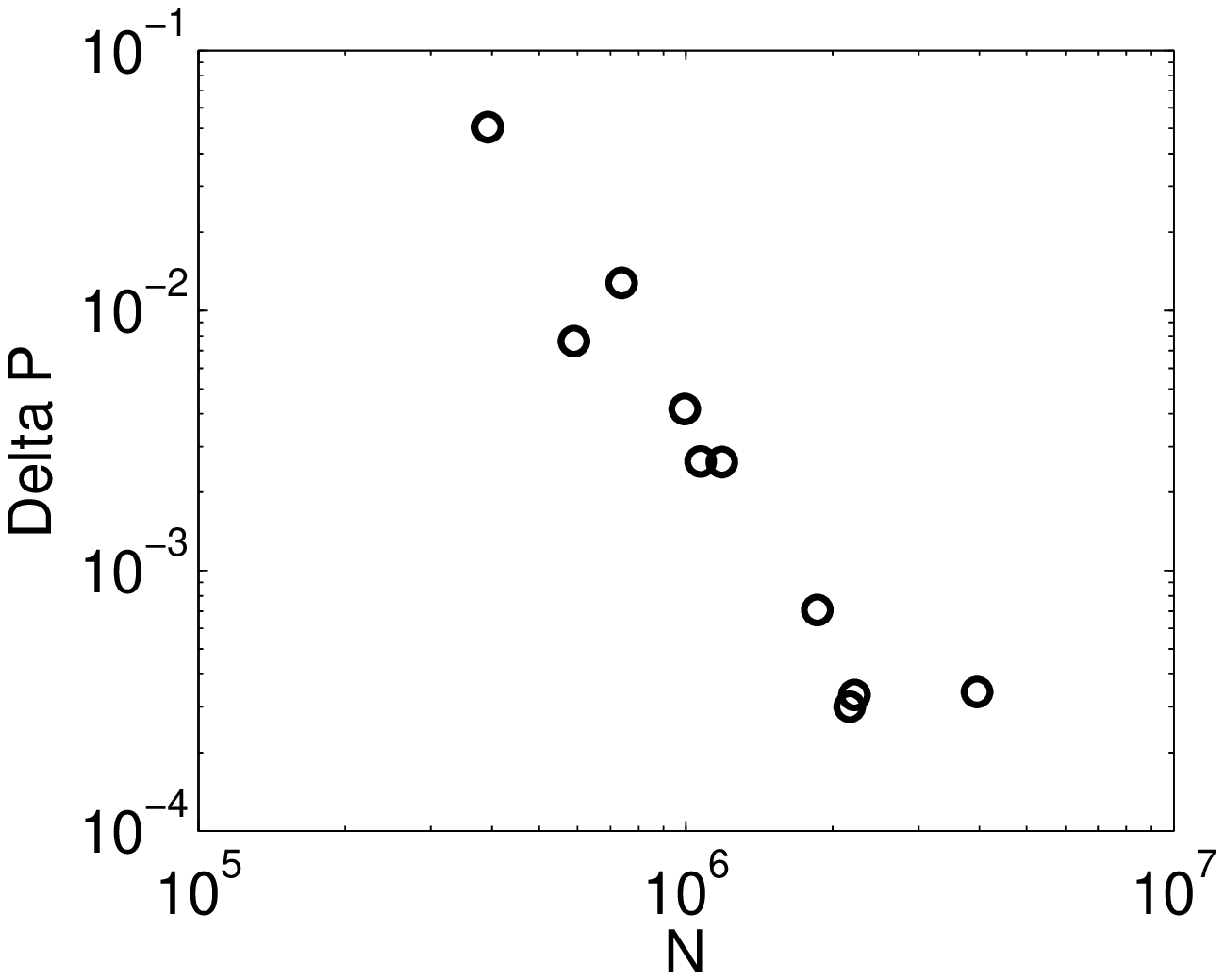}
  \caption{
From left to right: 
Mesh discretizing the 3D computational domain of a line mask. 
Relative error of the intensities of the zero, first, second and fifth diffraction orders, $|\vec{A}(\vec{k}_i)|^2$,
relative error of power conservation  $\Delta P$,
(cf.~Table~\ref{table_3d_full}).
}
\label{field_3d_full}
\end{center}
\end{figure}

First, the EUV line-mask is revisited using a 3D computational domain containing both, the absorber structure 
and the multi-layer mirror. 
Figure~\ref{field_3d_full} shows a mesh discretizing the geometry (generated automatically with the 
mesh generator JCMgeo). 
For the 3D setup the mesh consists of prismatoidal elements (instead of triangular elements as in the 2D 
setups). 
Simulations of the same physical setup have been performed using different spatial meshes with increasing mesh refinement 
and using 
finite-element ansatz-functions with varying polynomial degree $p$ (both, increasing $p$ and increasing mesh resolution 
in general leads to higher accuracy).  
The numerical results on intensities of several diffraction orders, on reflected power and on power conservation 
are given in Table~\ref{table_3d_full}.
From the results it can be seen that the first three significant digits of accuracy are reached for all investigated 
quantities. Considering the large computational domain with a size of the order of 10,000 cubic wavelengths 
this is a notable result which can be explained by the good convergence properties of higher-order 
finite-elements. 

Figure~\ref{field_3d_full} shows how the relative errors of the diffraction intensities converge with number of unknowns of the 
problem and how the power conservation error converges towards zero. 
As quasi-exact reference for the diffraction intensities, results from Sec.~\ref{sec_reference_1d_line_2ddd} are  used.

\subsubsection{1D line mask: 3D Domain decomposition results}
\label{sec_1d_line_3ddd}
\begin{figure}[t]
\begin{center}
  \includegraphics[width=.3\textwidth]{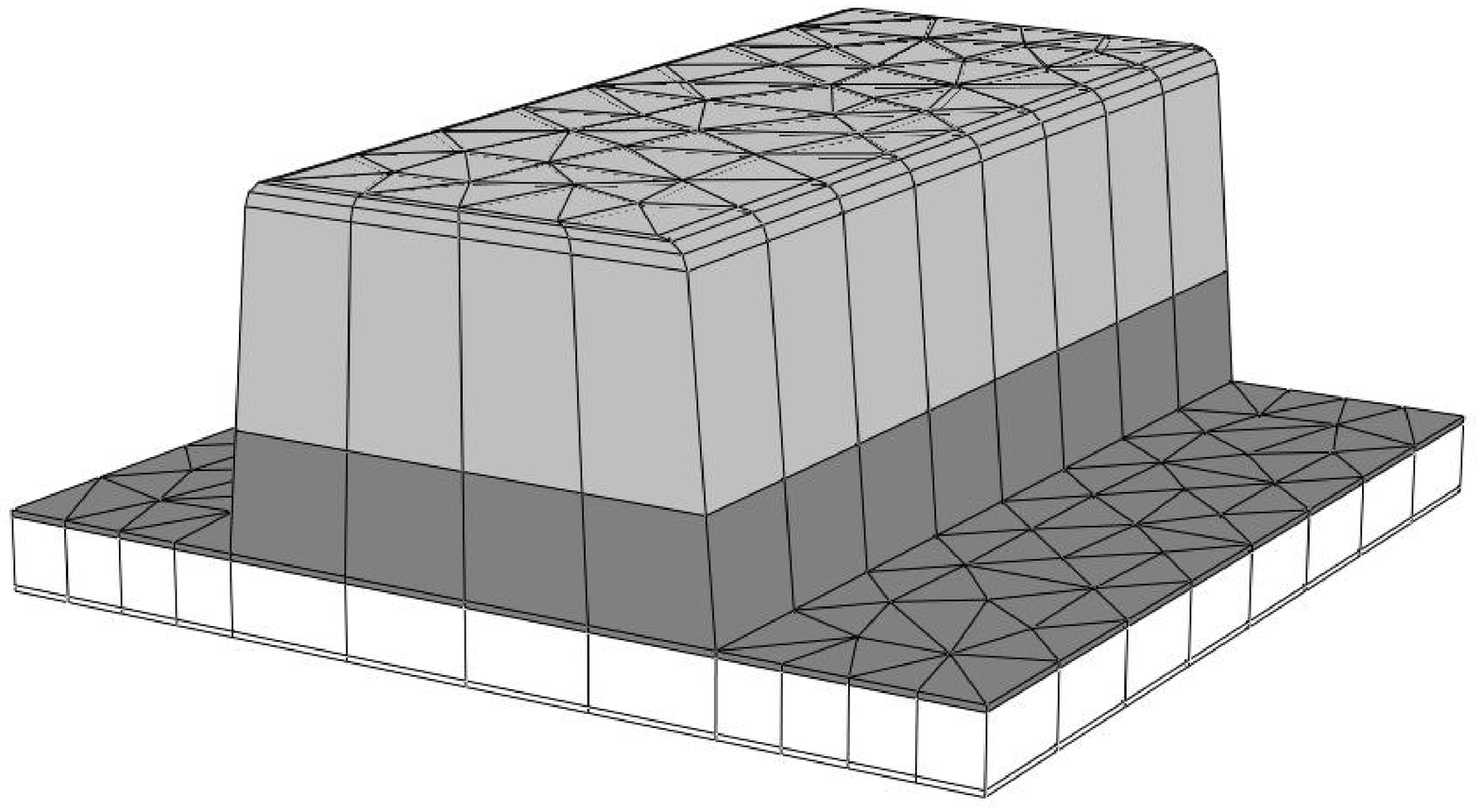}
  \hspace{.03\textwidth}
  \psfrag{N}{\sffamily $N$}
  \psfrag{Delta A}{\sffamily Abs.\,error}
  \includegraphics[width=.3\textwidth]{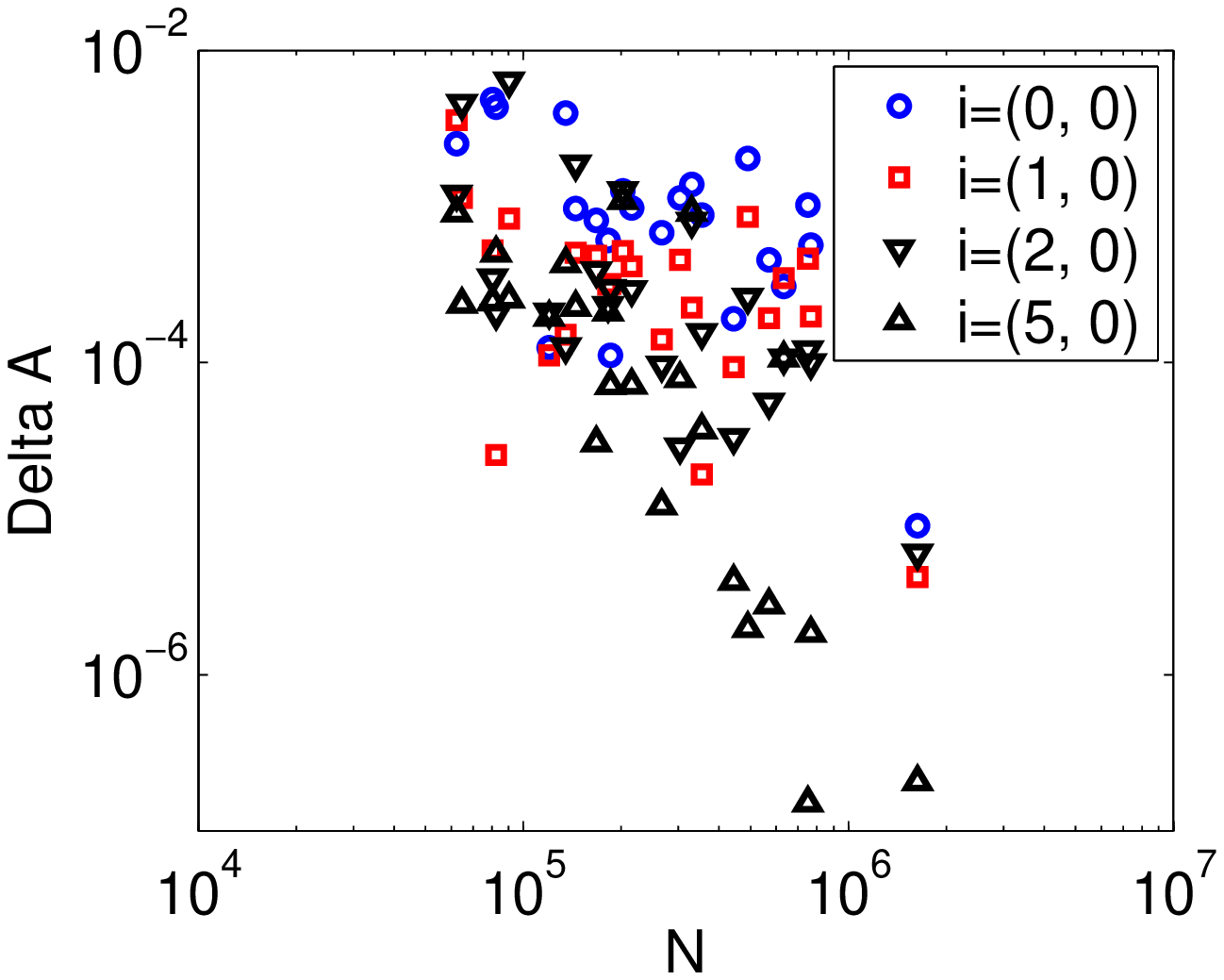}
  \hspace{.03\textwidth}
  \psfrag{Delta P}{\sffamily $(\Delta P_{\mbox{ref}})/P_{\mbox{ref}}$}
  \includegraphics[width=.3\textwidth]{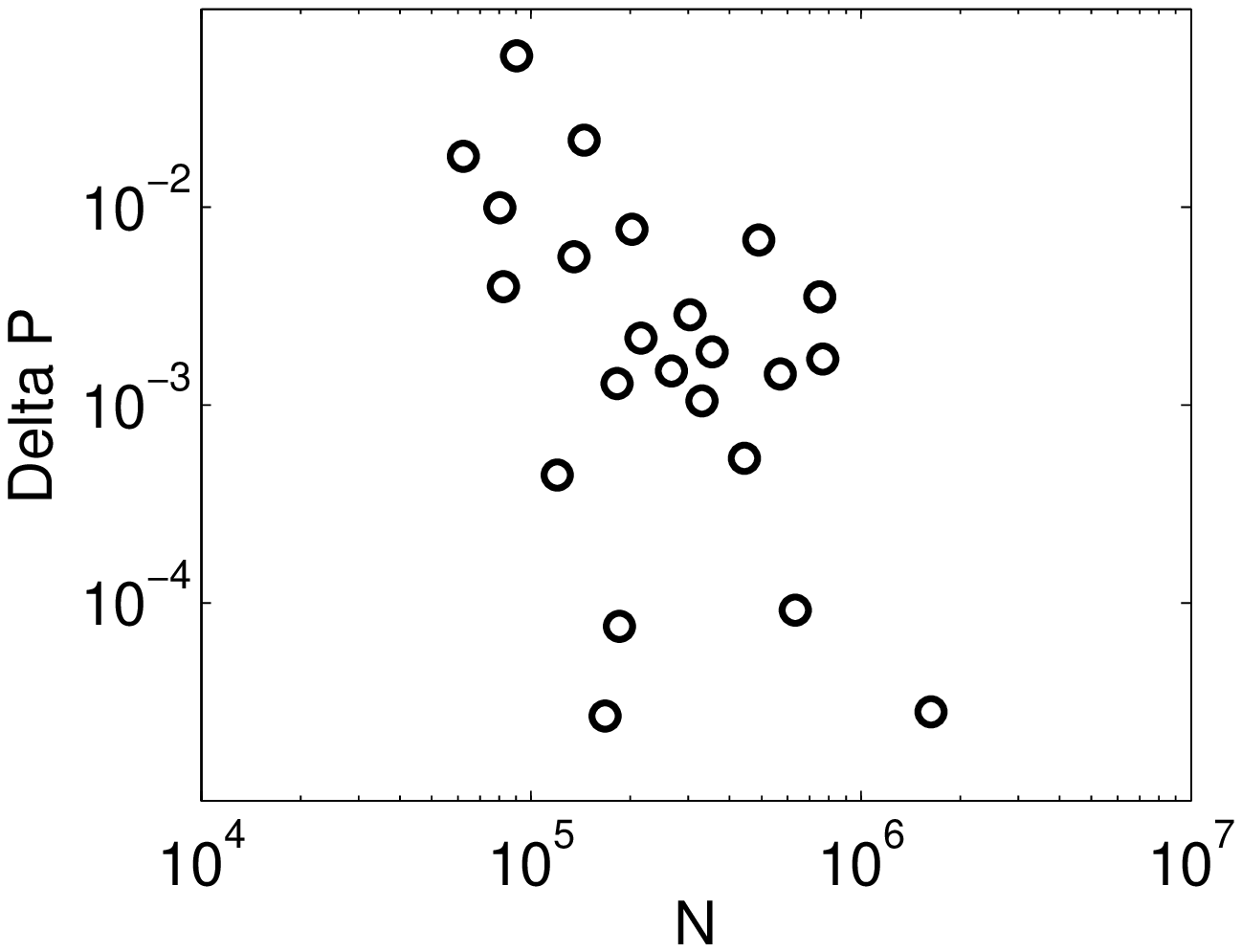}
  \caption{
From left to right: 
Mesh discretizing the 3D computational domain of a line mask (reduced computational domain for the 
domain-decomposition approach). 
Absolute errors of the intensities of the zero, first, second and fifth diffraction orders, $\Delta |\vec{A}(\vec{k}_i)|^2$,
relative error of reflected power  $\Delta P_{\mbox{ref}}$,
(cf.~Table~\ref{table_3d_dd}).
}
\label{field_3d_dd}
\end{center}
\end{figure}

In this Section simulation of the line-mask using a 3D computational domain and the domain-decomposition algorithm as described in 
Section~\ref{sec_reference_1d_line_2ddd} is demonstrated. 
Figure~\ref{field_3d_dd} shows a typical mesh discretizing the geometry. 
The numerical results are given in Table~\ref{table_3d_dd}. 
Figure~\ref{field_3d_dd} (center) shows the convergence of the (absolute) error of the diffraction intensities, 
$\Delta |\vec{A}(\vec{k}_i)|^2$ (for amplitudes $A_{S/P}=1$ of the incident light fields).
Accuracies in the range of $10^{-5}\dots 10^{-6}$ are reached. 
For the relative accuracy of the total reflected power, an accuracy below $10^{-4}$ is reached at highest numerical 
resolution. 
As quasi-exact reference for the diffraction intensities, results from Sec.~\ref{sec_reference_1d_line_2ddd} are  used. 
As can be seen from the table high numerical accuracy (with agreement of three to four significant digits even in 
low-power diffraction orders) is reached. 

\section{Simulations of 2D-periodic patterns on an EUV mask}
\label{sec_3d_simulations}

\begin{figure}[t]
\begin{center}
  \psfrag{px}{\sffamily $p_x$}
  \psfrag{py}{\sffamily $p_y$}
  \psfrag{CDx}{\sffamily \footnotesize $CD_{bottom,x}$}
  \psfrag{CDy}{\sffamily \footnotesize $CD_{bottom,y}$}
  \includegraphics[width=.3\textwidth]{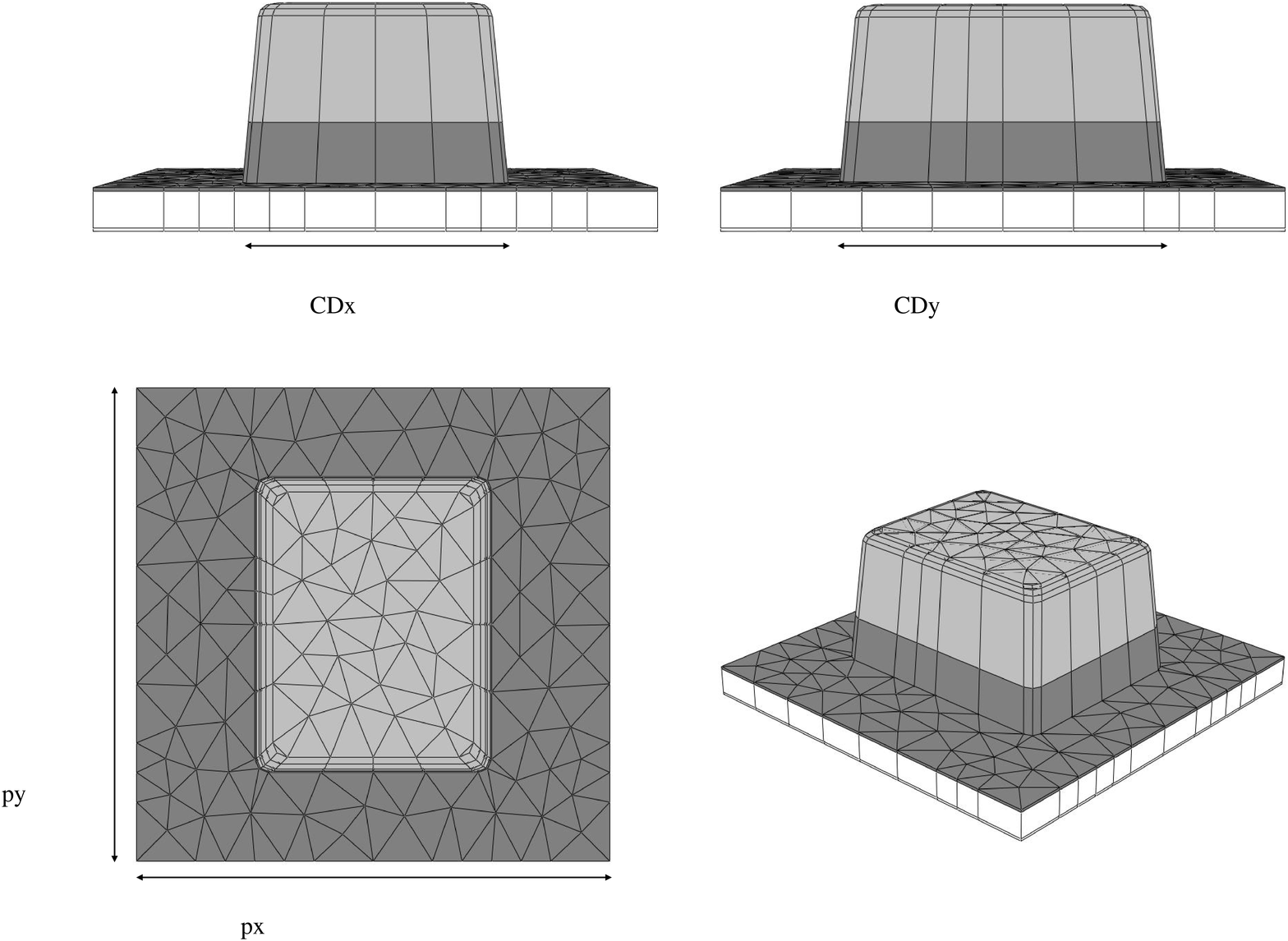}
  \hspace{.03\textwidth}
  \psfrag{Delta A}{\sffamily Rel.\,error}
  \includegraphics[width=.3\textwidth]{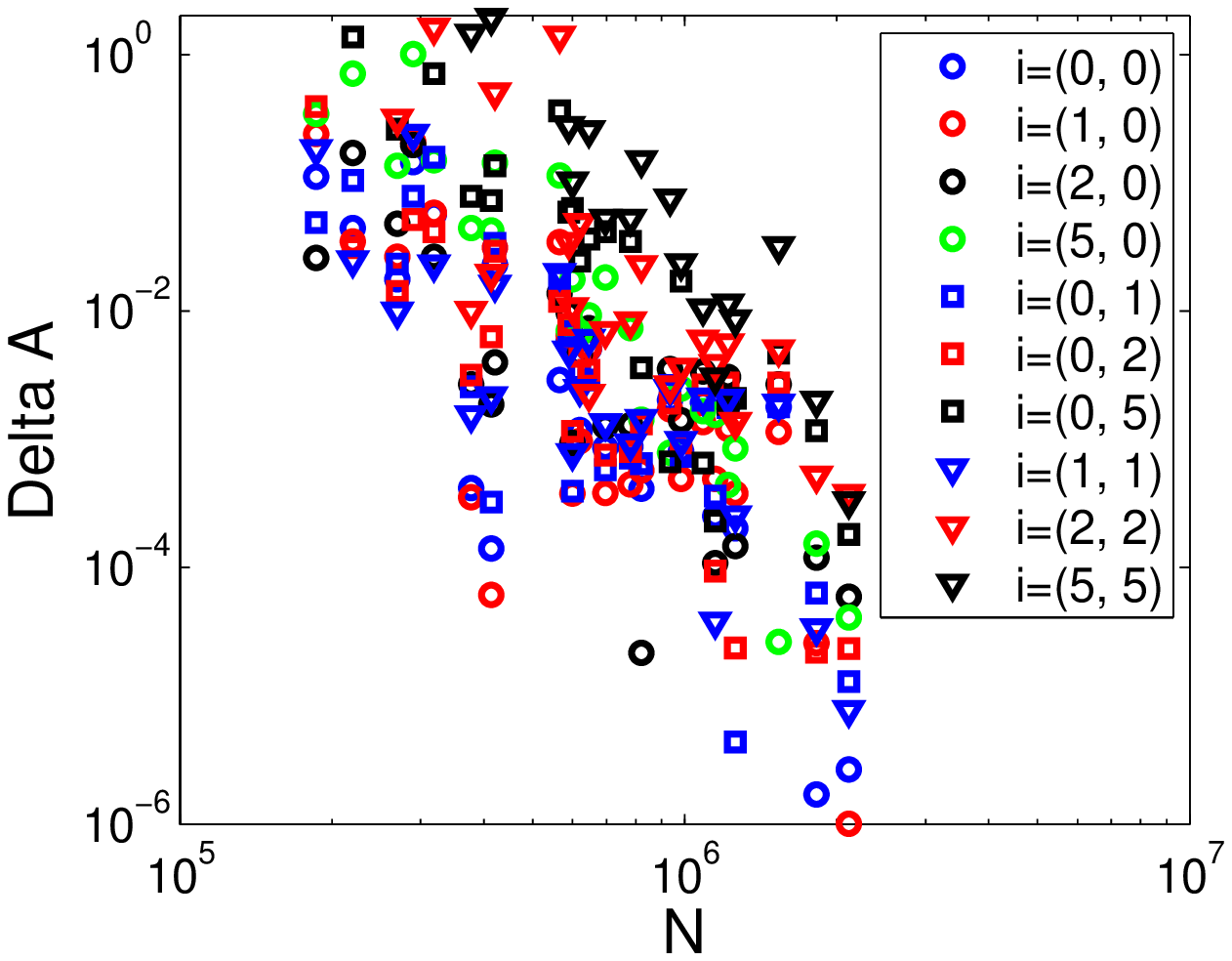}
  \hspace{.03\textwidth}
  \psfrag{Delta P}{\sffamily $(\Delta P_{\mbox{ref}})/P_{\mbox{ref}}$}
  \includegraphics[width=.3\textwidth]{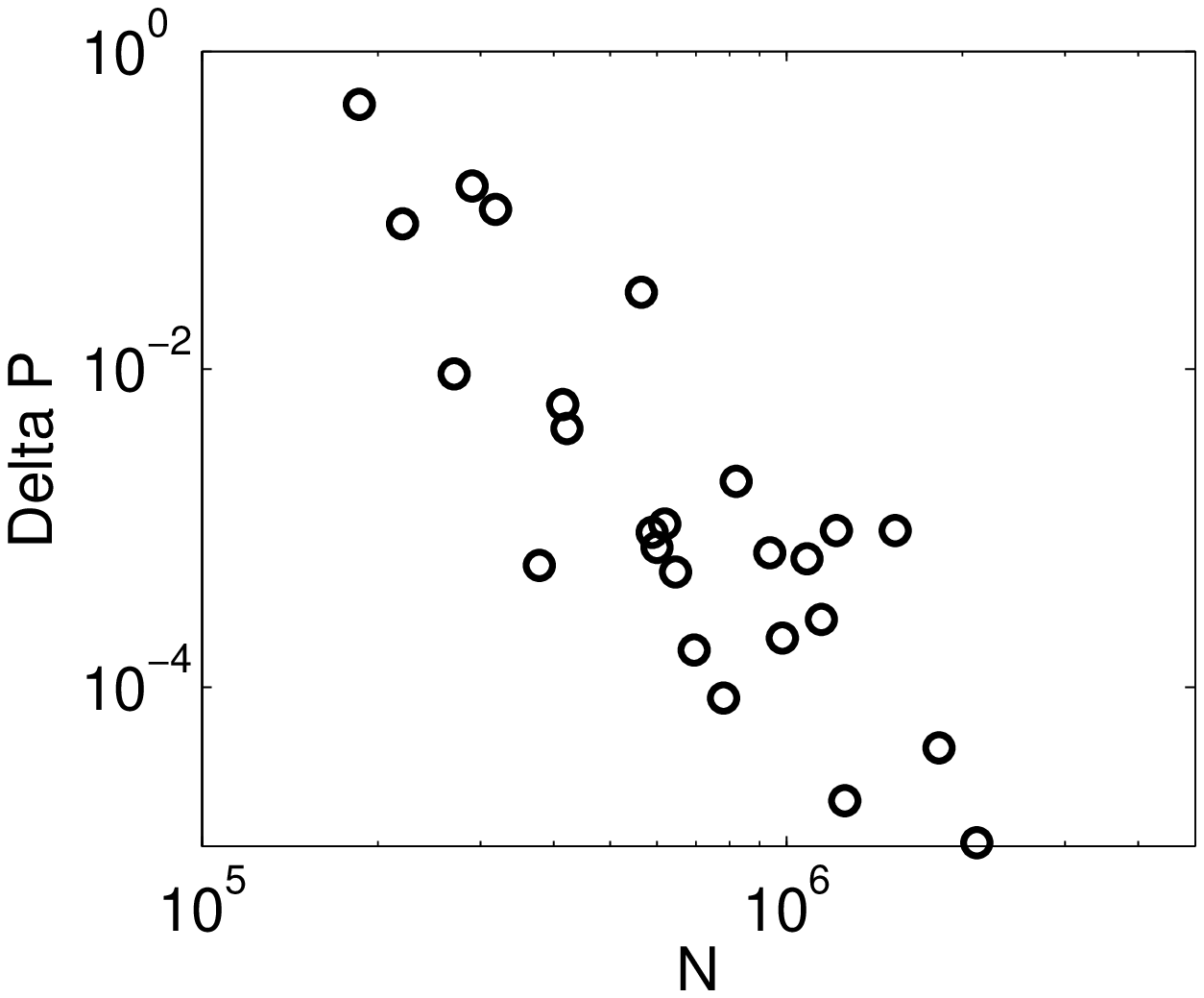}
  \caption{
Left: 
Mesh discretizing the 3D computational domain of a 2D periodic pattern (unit cell) on an EUV multi-layer mirror (top-view, side-views, and 
3D view). 
Right: Relative errors of the intensities of several diffraction orders, $|\vec{A}(\vec{k}_i)|^2$,
relative error of reflected power  $\Delta P_{\mbox{ref}}$, 
(cf.~Table~\ref{table_3d_block}).
}
\label{field_3d_block}
\end{center}
\end{figure}

In the previous Section it has been shown that the 3D light scattering solver module of JCMsuite 
generates numerical results which converge well to quasi-exact results obtained with the 2D light scattering 
module (which has been compared and benchmarked to independent rigorous methods and 
implementations~\cite{Burger2005bacus,Hoffmann2009a,Lockau2009a}). 
In this Section a 3D setup is investigated which cannot be reduced to a 2D computational domain: 
Figure~\ref{field_3d_block} shows the investigated setup. All parameters of the setup are detailed in 
Table~\ref{table_parameters}. 
The setup and execution of the simulations is performed as in Section~\ref{sec_1d_line_3ddd}.
Table~\ref{table_3d_block} presents simulation results for simulation on grids of different refinement levels and 
for finite-element ansatzfunctions of polynomial degree $p=3\dots 5$.
For this setup, also out-of-plane scattering takes place due to the 3D nature of the absorber block, therefore 
also some exemplary out-of-plane diffraction orders are included in the tabulation (e.g., diffraction order $A(\vec{k}_{1,1})$, 
first order diffraction in $x$- and in $y$-direction). 
As can be seen from the tabulated results and from the convergence of the numerical errors of the computed 
diffraction orders as displayed in Fig.~\ref{field_3d_block}, a high numerical accuracy is reached. 
Here, as quasi-exact result for the convergence plots, the FEM simulation result at highest numerical resolution 
has been chosen. 
The first three to four digits of the intensities even of relatively weak diffraction orders (at five orders of magnitude lower 
intensity than the most intense, zero diffraction order) are accurately 
computed.

\section{Conclusion}
Rigorous simulations of light scattering off 2D-periodic patterns on EUV masks have been performed. 
In a detailed convergence study it has been shown 
that high accuracies can be reached for the simulated intensities of the diffraction spectrum. 
This opens a prospect for scatterometric measurements of 3D patterns on EUV masks using FEM simulations for 
pattern reconstruction. 
Future work will concern application of reduced basis methods~\cite{pomplun:498} for significantly reducing computation times  
for this simulation task. This will open prospects for online reconstruction of 3D patterns on EUV masks. 

\section*{Acknowledgments}
The authors would like to acknowledge the support of
European Regional Development Fund (EFRE) / Investitionsbank Berlin (IBB) through contracts 
ProFIT 10144554 and 10144555.

\bibliography{/home/numerik/bzfburge/texte/biblios/phcbibli,/home/numerik/bzfburge/texte/biblios/group,/home/numerik/bzfburge/texte/biblios/lithography}
\bibliographystyle{spiebib}  

\vspace*{3cm}
\section*{Appendix: Tabulated simulation results}

\begin{table}[h]
{\footnotesize 
\begin{tabular}{|r|r|l|l|l|l|l|l|}
\hline
\input{tex_for_paper_2d_full.tex}
\hline
\end{tabular}}
\caption{
Simulation results for 2D simulations of an EUV line mask obtained on meshes with increasing resolution
(see Section~\ref{sec_reference_1d_line_2dfull}). 
Increased resolution corresponds to higher number of unknowns, $N$.  
Numerical results for magnitudes of several complex diffraction amplitudes (normalized to magnitude of incident light field), 
reflected power and power conservation 
converge with increasing resolution. Computation time $t$ in seconds (computations performed on a standard workstation). 
}
\label{table_2d_full}
\end{table}

\begin{table}[h]
{\footnotesize 
\begin{tabular}{|r|r|l|l|l|l|l|l|}
\hline
\input{tex_for_paper_2d_dd.tex}
\hline
\end{tabular}}
\caption{
Simulation results for 2D simulations of an EUV line mask obtained using a domain-decomposition algorithm
(see Section~\ref{sec_reference_1d_line_2ddd}). 
Results from simulations on the full 2D computational domain are given in the last data row for comparision. 
}
\label{table_2d_dd}
\end{table}

\begin{table}[h]
{\footnotesize 
\begin{tabular}{|r|r|r|l|l|l|l|l|l|l|}
\hline
\input{tex_for_paper_3d_full.tex}
\hline
\end{tabular}}
\caption{
Simulation results for 3D simulations of an EUV line mask. 
Results from simulations on 2D computational domains (reference solutions) are given in the last data rows for comparision
(see Section~\ref{sec_1d_line_3dfull}). 
}
\label{table_3d_full}
\end{table}

\begin{table}[h]
{\footnotesize 
\begin{tabular}{|r|r|r|l|l|l|l|l|l|}
\hline
\input{tex_for_paper_3d_dd.tex}
\hline
\end{tabular}}
\caption{
Simulation results for 3D simulations of an EUV line mask using the rigorous domain-decomposition approach. 
Results from simulations on 2D computational domains (reference solutions) are given in the last data rows for comparision
(see Section~\ref{sec_1d_line_3ddd}). 
}
\label{table_3d_dd}
\end{table}

\setlength{\tabcolsep}{0.5mm}
\begin{table}[h]
{\footnotesize 
\begin{tabular}{|r|r|r|l|l|l|l|l|l|l|l|l|l|}
\hline
\input{tex_for_paper_3d_block.tex}
\hline
\end{tabular}}
\caption{
Simulation results for 3D simulations of a 2D periodic absorber pattern on a multi-layer mirror 
(see Section~\ref{sec_3d_simulations}).
}
\label{table_3d_block}
\end{table}

\end{document}

%% file: tex_for_paper_2d_full.tex
N & $t$ [sec] & $|A(\vec{k}_0)|^2$ & $|A(\vec{k}_1)|^2$ & $|A(\vec{k}_2)|^2$ & $|A(\vec{k}_5)|^2$ & $P_{\mbox{ref}}/P_{\mbox{inc}} $ & $\Delta P$ \\ 
\hline 
  246686 &       25 &3.024847507e-01 & 1.906706398e-01 & 2.7206962e-02 & 1.0884158e-03 & 0.38217497 &4.01e-05 \\ 
  514148 &      192 &3.024385681e-01 & 1.906496392e-01 & 2.7206380e-02 & 1.0888107e-03 & 0.38212772 &1.01e-06 \\ 
 1110332 &      453 &3.024389867e-01 & 1.906489203e-01 & 2.7206608e-02 & 1.0884332e-03 & 0.38212733 &2.69e-09 \\ 
 2166904 &      955 &3.024389384e-01 & 1.906489036e-01 & 2.7206569e-02 & 1.0884331e-03 & 0.38212721 &7.16e-09 \\ 
 4115942 &     1489 &3.024389304e-01 & 1.906489032e-01 & 2.7206565e-02 & 1.0884326e-03 & 0.38212720 &5.24e-09 \\ 

%% file: tex_for_paper_2d_dd.tex
N & $t$ [sec] & $|A(\vec{k}_0)|^2$ & $|A(\vec{k}_1)|^2$ & $|A(\vec{k}_2)|^2$ & $|A(\vec{k}_5)|^2$ & $P_{\mbox{ref}}/P_{\mbox{inc}} $  \\ 
\hline 
   48490 &       13 & 3.024524519e-01 & 1.906621762e-01 & 2.7208305e-02 & 1.0875898e-03 & 0.38215086  \\ 
  119424 &       71 & 3.024373198e-01 & 1.906492090e-01 & 2.7207547e-02 & 1.0884615e-03 & 0.38212950  \\ 
  213726 &      161 & 3.024395380e-01 & 1.906493411e-01 & 2.7206633e-02 & 1.0884373e-03 & 0.38212813  \\ 
  447508 &      348 & 3.024390191e-01 & 1.906489541e-01 & 2.7206581e-02 & 1.0884328e-03 & 0.38212733  \\ 
  775440 &      519 & 3.024388938e-01 & 1.906488761e-01 & 2.7206563e-02 & 1.0884325e-03 & 0.38212716  \\ 
 1435086 &      840 & 3.024389134e-01 & 1.906488909e-01 & 2.7206563e-02 & 1.0884325e-03 & 0.38212718  \\ 
 2442526 &     1358 & 3.024389450e-01 & 1.906489122e-01 & 2.7206566e-02 & 1.0884325e-03 & 0.38212722  \\ 
 4439018 &     3049 & 3.024389436e-01 & 1.906489112e-01 & 2.7206566e-02 & 1.0884325e-03 & 0.38212722  \\ 
\hline 
\multicolumn{7}{|l|}{Results from Section~\ref{sec_reference_1d_line_2dfull} (2D, full computational domain):} \\ 
\hline 
 4115942 &     1489 &3.024389304e-01 & 1.906489032e-01 & 2.7206565e-02 & 1.0884326e-03 & 0.38212720  \\ 

%% file: tex_for_paper_3d_full.tex
N & p & $t$ [sec] &$|A(\vec{k}_0)|^2$ & $|A(\vec{k}_1)|^2$ & $|A(\vec{k}_2)|^2$ & $|A(\vec{k}_5)|^2$ & $P_{\mbox{ref}}/P_{\mbox{inc}} $ & $\Delta P$  \\ 
\hline 
  392600 &  4 &      188 &   0.30591 &   0.18950 & 0.02813 & 0.00049 &   0.38505 & 0.05064  \\ 
  588900 &  4 &      308 &   0.30271 &   0.19059 & 0.02718 & 0.00089 &   0.38199 & 0.00763  \\ 
  739200 &  5 &     1100 &   0.30367 &   0.19075 & 0.02739 & 0.00096 &   0.38280 & 0.01278  \\ 
  993720 &  4 &      625 &   0.30270 &   0.19091 & 0.02724 & 0.00110 &   0.38266 & 0.00419  \\ 
 1072800 &  5 &     1781 &   0.30292 &   0.19074 & 0.02730 & 0.00106 &   0.38278 & 0.00263  \\ 
 1184820 &  4 &      799 &   0.30283 &   0.19086 & 0.02727 & 0.00106 &   0.38270 & 0.00262  \\ 
 1859520 &  5 &     3951 &   0.30250 &   0.19060 & 0.02721 & 0.00109 &   0.38217 & 0.00070  \\ 
 2169440 &  4 &     4285 &   0.30257 &   0.19069 & 0.02723 & 0.00109 &   0.38229 & 0.00030  \\ 
 2217120 &  5 &     5209 &   0.30252 &   0.19059 & 0.02722 & 0.00108 &   0.38217 & 0.00033  \\ 
 3951360 &  5 &    21417 &   0.30269 &   0.19068 & 0.02728 & 0.00109 &   0.38237 & 0.00034  \\ 
\hline 
\multicolumn{9}{|l|}{Results from Section~\ref{sec_reference_1d_line_2dfull} (2D, full computational domain):} \\ 
\hline 
 4115942 &  4 &     1489 &   0.30244 &   0.19065 & 0.02721 & 0.00109 &   0.38213 & 5.24e-09  \\ 
\hline 
\multicolumn{9}{|l|}{Results from Section~\ref{sec_reference_1d_line_2ddd} (2D, domain decomposition):} \\ 
\hline 
 4439018 &  4 &     3049 &  0.30244 &   0.19065 & 0.02721 & 0.00109 &   0.38213  & \\ 

%% file: tex_for_paper_3d_dd.tex
N & p &$t$ [sec] & $|A(\vec{k}_0)|^2$ & $|A(\vec{k}_1)|^2$ & $|A(\vec{k}_2)|^2$ & $|A(\vec{k}_5)|^2$ & $P_{\mbox{ref}}/P_{\mbox{inc}}$   \\ 
\hline 
   64680 &  2 &       15 &   0.18196 &   0.18951 & 0.02266 & 0.00085 &   0.24844   \\ 
   90552 &  2 &       24 &   0.29084 &   0.18981 & 0.02087 & 0.00134 &   0.35996   \\ 
  144900 &  2 &       39 &   0.30147 &   0.19015 & 0.02535 & 0.00132 &   0.37385   \\ 
  202356 &  2 &       75 &   0.30118 &   0.19013 & 0.02595 & 0.00218 &   0.37917   \\ 
  330330 &  2 &      132 &   0.30382 &   0.19042 & 0.02641 & 0.00201 &   0.38173   \\ 
  631596 &  2 &      825 &   0.30275 &   0.19100 & 0.02731 & 0.00098 &   0.38209   \\ 
   62400 &  3 &       27 &   0.29991 &   0.18707 & 0.02601 & 0.00200 &   0.37524   \\ 
   82320 &  3 &       40 &   0.30677 &   0.19067 & 0.02700 & 0.00159 &   0.38364   \\ 
  135240 &  3 &       75 &   0.30639 &   0.19050 & 0.02708 & 0.00152 &   0.38427   \\ 
  182160 &  3 &      127 &   0.30304 &   0.19096 & 0.02744 & 0.00088 &   0.38262   \\ 
  303600 &  3 &      262 &   0.30358 &   0.19110 & 0.02724 & 0.00101 &   0.38322   \\ 
  568560 &  3 &      804 &   0.30289 &   0.19084 & 0.02726 & 0.00109 &   0.38267   \\ 
   80600 &  4 &       64 &   0.30729 &   0.19013 & 0.02755 & 0.00084 &   0.38590   \\ 
  120120 &  4 &      194 &   0.30256 &   0.19076 & 0.02742 & 0.00089 &   0.38196   \\ 
  185380 &  4 &      206 &   0.30255 &   0.19097 & 0.02750 & 0.00116 &   0.38216   \\ 
  265980 &  4 &      351 &   0.30311 &   0.19079 & 0.02730 & 0.00108 &   0.38269   \\ 
  443300 &  4 &      714 &   0.30263 &   0.19074 & 0.02724 & 0.00108 &   0.38233   \\ 
  749840 &  4 &     3775 &   0.30346 &   0.19111 & 0.02733 & 0.00109 &   0.38347   \\ 
  168000 &  5 &      422 &   0.30325 &   0.19113 & 0.02759 & 0.00106 &   0.38212   \\ 
  215040 &  5 &      296 &   0.30342 &   0.19106 & 0.02750 & 0.00102 &   0.38296   \\ 
  353280 &  5 &      601 &   0.30332 &   0.19067 & 0.02736 & 0.00105 &   0.38283   \\ 
  491040 &  5 &     2007 &   0.30446 &   0.19150 & 0.02747 & 0.00109 &   0.38473   \\ 
  765600 &  5 &     2070 &   0.30300 &   0.19085 & 0.02731 & 0.00109 &   0.38278   \\ 
 1631520 &  5 &    16760 &   0.30245 &   0.19065 & 0.02721 & 0.00109 &   0.38214   \\ 
\hline 
\multicolumn{8}{|l|}{Results from Section~\ref{sec_reference_1d_line_2dfull} (2D, full computational domain):} \\ 
\hline 
 4115942 & &     1489 &   0.30244 &   0.19065 & 0.02721 & 0.00109 &   0.38213   \\ 
\hline 
\multicolumn{8}{|l|}{Results from Section~\ref{sec_reference_1d_line_2ddd} (2D, domain decomposition):} \\ 
\hline 
 4439018 & &     3049 &  0.30244 &   0.19065 & 0.02721 & 0.00109 &   0.38213   \\ 
\hline 
\multicolumn{8}{|l|}{Results from Section~\ref{sec_1d_line_3dfull} (3D, full computational domain):} \\ 
\hline 
 3951360 &     5 &    21417 &  0.30269 &   0.19068 & 0.02728 & 0.00109 &   0.38237   \\ 

%% file: tex_for_paper_3d_block.tex
N & p &$t$ [s] & $|A(\vec{k}_{0,0})|^2$ & $|A(\vec{k}_{1,0})|^2$  & $|A(\vec{k}_{2,0})|^2$   & $|A(\vec{k}_{5,0})|^2$ & $|A(\vec{k}_{0,1})|^2$  & $|A(\vec{k}_{0,2})|^2$  & $|A(\vec{k}_{0,5})|^2$ & $|A(\vec{k}_{1,1})|^2$ &  $|A(\vec{k}_{5,5})|^2$  & {\tiny $P_{\mbox{ref}}/P_{\mbox{inc}}$}   \\ 
\hline 
  186000 &  3 &      221 & 0.591897 & 0.051198 & 0.006972 & 4.683e-04 & 0.040457 & 0.009346 & 4.323e-04 & 0.016145 & 2.503e-04 & 0.48871   \\ 
  220320 &  3 &      405 & 0.636367 & 0.064991 & 0.008376 & 5.969e-04 & 0.038073 & 0.006505 & 6.007e-05 & 0.019269 & 6.676e-05 & 0.50070   \\ 
  270000 &  3 &      332 & 0.654288 & 0.065556 & 0.007503 & 3.012e-04 & 0.041543 & 0.006620 & 1.874e-05 & 0.019562 & 1.715e-05 & 0.50366   \\ 
  378000 &  3 &      457 & 0.665727 & 0.067312 & 0.007178 & 3.332e-04 & 0.042424 & 0.006694 & 2.738e-05 & 0.019788 & 9.737e-07 & 0.51085   \\ 
  414000 &  3 &      530 & 0.665910 & 0.067331 & 0.007145 & 3.339e-04 & 0.042520 & 0.006673 & 2.723e-05 & 0.019715 & 1.161e-06 & 0.51094   \\ 
  600000 &  3 &      776 & 0.666584 & 0.067361 & 0.007165 & 3.426e-04 & 0.042550 & 0.006723 & 2.699e-05 & 0.019773 & 4.375e-07 & 0.51150   \\ 
  696000 &  3 &      949 & 0.666577 & 0.067361 & 0.007168 & 3.423e-04 & 0.042558 & 0.006720 & 2.644e-05 & 0.019784 & 4.179e-07 & 0.51152   \\ 
  780000 &  3 &     1160 & 0.666587 & 0.067365 & 0.007168 & 3.461e-04 & 0.042564 & 0.006720 & 2.628e-05 & 0.019776 & 4.178e-07 & 0.51154   \\ 
  290160 &  4 &      734 & 0.762650 & 0.081339 & 0.008564 & 7.009e-04 & 0.039188 & 0.007061 & 1.569e-04 & 0.024497 & 1.821e-04 & 0.61192   \\ 
  318240 &  4 &      744 & 0.704052 & 0.071249 & 0.007349 & 2.963e-04 & 0.049282 & 0.006995 & 4.334e-05 & 0.019305 & 3.055e-05 & 0.54827   \\ 
  421200 &  4 &      867 & 0.681309 & 0.069445 & 0.007187 & 3.988e-04 & 0.043970 & 0.006913 & 2.195e-05 & 0.020076 & 4.938e-06 & 0.52459   \\ 
  589680 &  4 &     1139 & 0.669892 & 0.067774 & 0.007228 & 3.463e-04 & 0.042828 & 0.006768 & 2.389e-05 & 0.019855 & 5.069e-07 & 0.51424   \\ 
  645840 &  4 &     1273 & 0.669373 & 0.067682 & 0.007212 & 3.455e-04 & 0.042657 & 0.006738 & 2.446e-05 & 0.019877 & 4.983e-07 & 0.51378   \\ 
  936000 &  4 &     2157 & 0.667351 & 0.067451 & 0.007184 & 3.484e-04 & 0.042645 & 0.006728 & 2.541e-05 & 0.019807 & 4.269e-07 & 0.51224   \\ 
 1085760 &  4 &     2942 & 0.667295 & 0.067427 & 0.007183 & 3.481e-04 & 0.042619 & 0.006733 & 2.541e-05 & 0.019799 & 3.929e-07 & 0.51215   \\ 
 1216800 &  4 &     3811 & 0.667244 & 0.067417 & 0.007181 & 3.486e-04 & 0.042609 & 0.006733 & 2.534e-05 & 0.019799 & 4.016e-07 & 0.51211   \\ 
  565440 &  5 &     1816 & 0.667937 & 0.069668 & 0.007257 & 3.089e-04 & 0.041770 & 0.006795 & 3.462e-05 & 0.020149 & 9.692e-06 & 0.51395   \\ 
  620160 &  5 &     1901 & 0.665217 & 0.067270 & 0.007191 & 3.464e-04 & 0.042390 & 0.006685 & 2.477e-05 & 0.019709 & 1.296e-06 & 0.51047   \\ 
  820800 &  5 &     2604 & 0.665730 & 0.067297 & 0.007159 & 3.492e-04 & 0.042506 & 0.006724 & 2.548e-05 & 0.019729 & 4.563e-07 & 0.51088   \\ 
 1149120 &  5 &     3977 & 0.665837 & 0.067302 & 0.007158 & 3.482e-04 & 0.042518 & 0.006714 & 2.538e-05 & 0.019757 & 3.959e-07 & 0.51097   \\ 
 1258560 &  5 &     4833 & 0.665870 & 0.067310 & 0.007158 & 3.484e-04 & 0.042533 & 0.006715 & 2.544e-05 & 0.019753 & 4.005e-07 & 0.51100   \\ 
 1824000 &  5 &     8336 & 0.666003 & 0.067334 & 0.007158 & 3.487e-04 & 0.042531 & 0.006715 & 2.542e-05 & 0.019757 & 3.978e-07 & 0.51112   \\ 
 2115840 &  5 &    11109 & 0.666003 & 0.067336 & 0.007158 & 3.487e-04 & 0.042533 & 0.006715 & 2.538e-05 & 0.019758 & 3.969e-07 & 0.51112   \\ 
 2371200 &  5 &    14312 & 0.666004 & 0.067335 & 0.007159 & 3.487e-04 & 0.042534 & 0.006715 & 2.539e-05 & 0.019758 & 3.971e-07 & 0.51113   \\ 
\hline 